\documentclass[runningheads,fleqn]{svmult}
\usepackage{makeidx}   
\usepackage{graphicx}  
\usepackage{subeqnar}  
\usepackage{multicol}  
\usepackage{taphys}    
\usepackage{latexsym}
\makeindex             
\begin{document}
\title*{Real-Space Renormalization and Energy-Level Statistics at the
  Quantum Hall Transition}
\toctitle{Real-space renormalization and energy-level statistics
\protect\newline at the quantum Hall transition}
%
%
\titlerunning{Real-space renormalization at the quantum Hall transition}
%
\author{Rudolf A.\ R\"{o}mer\inst{1}
\and Philipp Cain\inst{2}}
\authorrunning{Rudolf A.\ R\"{o}mer and Philipp Cain}
%
%
\institute{Department of Physics and Centre for Scientific Computing, University of Warwick, Coventry, CV4 7AL, United Kingdom
\and Institut f\"{u}r Physik, Technische Universit\"{a}t Chemnitz, 09107 Chemnitz, Germany}

\maketitle              

\begin{abstract}
  We review recent applications of the real-space renormalization group
  (RG) approach to the integer quantum Hall (QH) transition.  The RG
  approach, applied to the Chalker-Coddington network model, reproduces
  the critical distribution of the {\em power} transmission
  coefficients, i.e., two-terminal conductances, $P_{\rm c}(G)$, with
  very high accuracy. The RG flow of $P(G)$ at energies away from the
  transition yields a value of the critical exponent,
  $\nu_{G}=2.39\pm0.01$, that agrees with most accurate large-size
  lattice simulations.  Analyzing the evolution of the distribution of
  {\em phases} of the transmission coefficients upon a step of the RG
  transformation, we obtain information about the energy-level
  statistics (ELS). From the fixed point of the RG transformation we
  extract a critical ELS.  Away from the transition the ELS crosses over
  towards a Poisson distribution.  Studying the scaling behavior of the
  ELS around the QH transition, we extract the critical exponent
  $\nu_{\rm ELS}=2.37\pm0.02$.
\end{abstract}

The integer quantum Hall (QH) transition is described well in terms of a
delocalization-localization transition of the electronic wavefunctions.
In contrast to a usual metal-insulator transition (MIT), the QH
transition is characterized by a single extended state located exactly
at the center $\epsilon=0$ of each Landau band \cite{Huc95}.  When
approaching $\epsilon=0$, the localization length $\xi$ of the electron
wavefunction diverges according to a power law $\epsilon^{-\nu}$, where
$\epsilon$ defines the distance to the MIT for a suitable control
parameter, e.g., the electron energy. On the theoretical side, the value
of $\nu$ has been extracted from various numerical simulations, e.g.,
$\nu=2.5\pm0.5$ \cite{ChaC88}, $2.4\pm0.2$ \cite{LeeWK93},
$2.35\pm0.03$ \cite{Huc92}, and $2.39\pm0.01$ \cite{CaiRSR01}.  In
experiments $\nu\approx 2.3$ has been obtained, e.g., from the frequency
\cite{HohZH01a} or the sample size \cite{KocHKP91b} dependence of the
critical behavior of the resistance in the transition region at strong
magnetic field.

Recently, a semianalytical description of the integer QH
transition, based on the extension of the scaling ideas for the
classical percolation \cite{StaA92} to the Chalker-Coddington (CC)
model of the quantum percolation \cite{ChaC88}, has been developed
\cite{GalR97,AroJS97}. The key idea of this description, a
real-space-renormalization group (RG) approach, is the following.
Each RG step corresponds to a doubling of the system size. The RG
transformation relates the conductance {\em distribution} of the
sample at the next step to the conductance distribution at the
previous step.  The {\em fixed point} (FP) of this transformation,
yields the distribution of the conductance, $P_{\rm c}(G)$, of a
{\em macroscopic} sample at the QH transition. This {\em
  universal} distribution describes the mesoscopic properties of a
fully coherent QH sample.  Analogously to the classical
percolation \cite{StaA92}, the correlation length exponent, $\nu$, was
extracted from the RG procedure \cite{CaiRSR01} using the fact that a
slight shift of the initial distribution with respect to the
FP distribution $P_{\rm c}(G)$ drives the system to the insulator upon
renormalization. Then the rate of the shift of the distribution
maximum determines the value of $\nu$.  Remarkably, both $P_{\rm
  c}(G)$ and the critical exponent obtained within the RG approach
\cite{CaiRSR01,WeyJ98,JanMMW98} agree very well with the ``exact''
results of the large-scale simulations
\cite{LeeWK93,Huc92,WanJL96,WanLS98,AviBB99}.

The goal of the present paper is threefold.  First, we briefly
review the basic ingredients that constitute the real-space RG
method in the QH situation \cite{CaiRSR01}.
Second, we extend the RG approach to include the level statistics
at the QH transition and apply a method analogous to the
finite-size-corrections analysis to extract $\nu$ from the
energy-level statistics (ELS) obtained within the RG approach.
This method yields $\nu = 2.37 \pm 0.02$, which is even closer to
the most precise large-scale simulations result $\nu = 2.35 \pm
0.03$ \cite{Huc92} than the value $\nu = 2.39 \pm 0.01$ inferred
from the conductance distribution \cite{CaiRSR01}. This agreement
is by no means trivial. Indeed, the original RG transformation
\cite{CaiRSR01} related the conductances, i.e., the {\em absolute
values} of the transmission coefficients of the original and the
doubled samples, while the {\em
  phases} of the transmission coefficients were assumed random and
uncorrelated.  In contrast, the level statistics at the transition
corresponds to the FP in the distribution of these phases.
Therefore, the success of the RG approach for conductances does not
guarantee that it will be equally accurate {\em quantitatively} for the
level statistics.
Third, we show that the RG structure employed in the present approach,
which is constructed from $5$ saddle points (SP), represents in many
aspects the minimal model of the QH transition. A further reduction in
the number of SP leads to less reliable results.

\section{The RG approach to the CC model}
\label{sec-rg-result}

\begin{figure}[tb]
\centerline{\includegraphics[width=0.4\columnwidth]{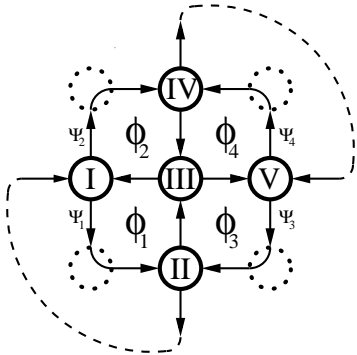}}
\caption{\label{fig-lsd-RGstruct} \label{fig-RGstruct}
  Chalker-Coddington network on a square lattice consisting of nodes
  (circles) and links (arrows). The RG unit used to construct the
  matrix (\ref{eq-lsd-system}) combines five nodes (full circles) by
  neglecting some connectivity (dashed circles).  $\Phi_1, \ldots,
  \Phi_4$ are the phases acquired by an electron along the loops as
  indicated by the arrows. $\Psi_1, \ldots, \Psi_4$ represent wave
  function amplitudes, and the thin dashed lines illustrate the
  boundary conditions used for the computation of level statistics}
\end{figure}

\begin{figure}[t]
\begin{center}
\includegraphics[width=0.7\columnwidth]{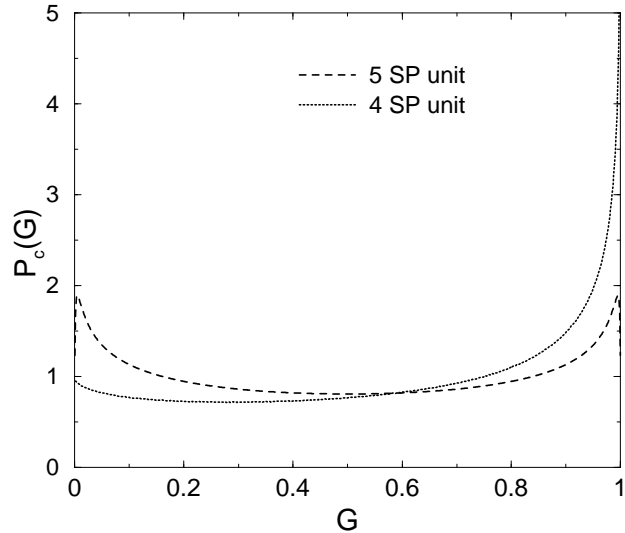}
\caption{\label{fig-rg-pgcomp}
  The critical distribution of the conductance $P_{\rm c}(G)$ at the QH
  transition obtained using the $5$ SP RG unit (dashed line). The dotted line
  denotes a $4$ SP RG unit as discussed in Sect.\
  \protect\ref{sec-rg-diff}. The latter distribution clearly deviates
  from the expected symmetry with respect to $G=0.5$  }
\end{center}
\end{figure}

Our RG approach to the QH transition \cite{CaiRSR01,GalR97,AroJS97} is
based on the RG unit shown in Fig.\ \ref{fig-RGstruct}.
The unit is a fragment of the CC network consisting of five nodes.  Each
node, $i$, is characterized by the transmission coefficient $t_i$, which
is an amplitude to deflect an incoming electron along the link to the
left.  Analogously, the reflection coefficient $r_i = (1-t_i^2)^{1/2}$
is the amplitude to deflect the incoming electron to the right.
Doubling of the sample size corresponds to the replacement of the RG
unit by a single node.  The RG transformation expresses the transmission
coefficient of this effective node, $t^{\prime}$, through the
transmission coefficients of the five constituting nodes \cite{GalR97}
\begin{equation}
  \label{eq-qhrg}
  t'= \left | \frac{
  \begin{array}{lr}
  t_1 t_5 \left(r_2 r_3 r_4 e^{i \Phi_2} - 1\right) +
  & t_2
      t_4 e^{i  (\Phi_3+\Phi_4)} \left(r_1 r_3 r_5 e^{-i \Phi_1} -
      1\right) \mbox{ \quad} \\[0.5ex]
      & + t_3 \left(t_2 t_5 e^{i \Phi_3} + t_1 t_4 e^{i \Phi_4}\right)
      \end{array} }
    { \left(r_3 - r_2 r_4 e^{i \Phi_2}\right) \left(r_3 - r_1 r_5 e^{i
        \Phi_1}\right) + \left(t_3 - t_4 t_5 e^{i \Phi_4}\right) \left(t_3 - t_1 t_2
      e^{i \Phi_3}\right) }\right |.
\end{equation}
Here $\Phi_j$ are the phases accumulated along the closed loops
(Fig.\ \ref{fig-RGstruct}). Within the RG approach to the
conductance distribution, information about electron energy is
incorporated only into the values of $t_i$ \cite{CaiRSR01}. The
energy dependence of phases, $\Phi_j$, is irrelevant; they are
assumed completely random. Due to this randomness, the
transmission coefficients, $t_i$, for a given energy, are also
randomly distributed with a distribution function $P(t)$.  Then
the transformation (\ref{eq-qhrg}) allows, upon averaging over
$\Phi_j$, to generate the next-step distribution $P(t^{\prime})$.
Therefore, within the RG scheme, a delocalized state corresponds
to the FP distribution $P_{\rm c}(t)$ of the RG transformation.
Due to the symmetry of the RG unit, it is obvious that the
critical distribution, $P_{\rm c}(t^2)$, of the power transmission
coefficient, $t^2=G$, which has the meaning of the two-terminal
conductance, is symmetric with respect to $t^2=\frac{1}{{2}}$ as
shown in Fig.\ \ref{fig-rg-pgcomp}.
In other words, the RG transformation respects the duality between
transmission and reflection.  The critical distribution $P_{\rm c}(G)$
found in Refs.\ \cite{CaiRSR01} and \cite{GalR97} agrees very well with
the results of direct large-scale simulations.

\section{RG Approach to the ELS}

It has been realized long ago that, alongside with the change in the
behavior of the eigenfunctions, a localization-delocalization transition
manifests itself in the statistics of the energy levels. In particular,
as the energy is swept across the mobility edge, the shape of the ELS
crosses over from the Wigner-Dyson distribution, corresponding to the
appropriate universality class, to the Poisson distribution.
Moreover, finite-size corrections to the critical ELS close to the
mobility edge allow to determine the value of the correlation
length exponent \cite{ShkSSL93}, thus avoiding an actual analysis
of the spatial extent of the wave functions.  For this reason, the
ELS constitutes an alternative to the MacKinnon-Kramer
\cite{PicS81a,PicS81b,MacK81,MacK83} and to the
transmission-matrix \cite{Lan70,FisL81} approaches to the
numerical study of localization.

\subsection{Derivation of the Network Operator for the RG}

As has been shown by Fertig \cite{Fer88}, energy levels of an 2D CC
network can be computed from the energy dependence of the so called
network operator $U(E)$.
$U$ is constructed similar to the system of equations for
obtaining the transmission coefficient $t'$ of the RG unit as
presented in Eq.\ (\ref{eq-qhrg}).  Every SP of the network
contributes two scattering equations. Each of them describes the
amplitude of one outgoing channel using the amplitudes of the two
incoming channels weighted by the transmission and reflection
coefficients $t$ and $r$ in which also the random phase $\Phi$ of
the links between SP's can be incorporated. When comparing to the
calculation of the transmission coefficient $t'$ an essential
difference has to be taken into account. Energy levels are defined
only in a closed system which requires to apply appropriate
boundary conditions.
The energy dependence of $U(E)$ enters trough the energy
dependence of the $t_i(E)$ of the SP's, as well as the energy
dependence of the phases $\Phi_j(E)$ of the links.  Considering
the vector $\Psi$ of wave amplitudes on the links of the network,
the eigenenergies can now be obtained from the stationary
condition
\begin{equation}
\label{eq-lsd-statU}
U(E) \Psi = \Psi .
\end{equation}
Nontrivial solutions exist only for discrete energies $E_k$, which
coincide with the eigenenergies of the system \cite{Fer88}. The
evaluation of the $E_k$'s according to Eq.\ (\ref{eq-lsd-statU}) is
numerically very expensive. For that reason a simplification was
proposed \cite{KleM97}. Instead of solving the real eigenvalue problem,
calculating a spectrum of quasienergies $\omega$ is suggested
following from
\begin{equation}
\label{eq-lsd-statUomega}
U(E) \Psi_l = e^{i  \omega_l(E)} \Psi_l .
\end{equation}
For fixed energy $E$ the $\omega_l$ are expected to obey the same
statistics as the real eigenenergies \cite{KleM97}. This approach makes
is perfectly suited for large-size numerical simulations, e.g.
studying $50\times50$ SP networks.

In order to combine the above algorithm with the RG iteration, in which
a rather small unit of SP's is considered, we first ``close'' the RG
unit at each RG step in order to discretize the energy levels as shown
in Fig.\ \ref{fig-lsd-RGstruct} with dashed lines.
For a given closed RG unit with a fixed set of $t_i$-values at the
nodes, the positions of the energy levels are determined by the
energy dependences, $\Phi_j(E)$, of the four phases along the
loops. These phases change by $\sim \pi$ within a very narrow
energy interval, inversely proportional to the sample size. Within
this interval the change of the transmission coefficients is
negligibly small. A closed RG unit in Fig.\ \ref{fig-lsd-RGstruct}
contains $10$ links, and, thus, it is described by $10$
amplitudes. Each link is characterized by an individual phase. On
the other hand, it is obvious that the energy levels are
determined only by the phases along the loops. One way to derive
$U$ is to combine the individual phases into phases $\Phi_j$
connected to the four inner loops of the unit and to exclude from
the original system of $10$ equations all amplitudes except the
``boundary'' amplitudes $\Psi_j$ (Fig.\ \ref{fig-lsd-RGstruct}).
The network operator for the remaining four amplitudes is a
$4\times 4$ matrix $\{U_{nm}\}$ with elements
\begin{equation}
\begin{array}{ll}
\label{eq-lsd-system}
U_{11} =  (r_1 r_2 - t_1 t_2 t_3) e^{-i \Phi_1} \quad
&U_{12} =  (t_1 r_2 + t_2 t_3 r_1) e^{-i \Phi_1} \nonumber \\
U_{13} =  t_2 t_5 r_3 e^{-i \Phi_1}  \quad
&U_{14} =  t_2 r_3 r_5 e^{-i \Phi_1} \nonumber \\[1ex]
U_{21} =  -t_1 r_3 r_4 e^{-i \Phi_2}  \quad
&U_{22} =  r_1 r_3 r_4 e^{-i \Phi_2} \nonumber \\
U_{23} =  -(t_4 r_5 + t_3 t_5 r_4) e^{-i \Phi_2}  \quad
&U_{24} =  (t_4 t_5 - t_3 r_4 r_5) e^{-i \Phi_2}\nonumber \\[1ex]
U_{31} =  -t_1 t_4 r_3 e^{-i \Phi_4}  \quad
&U_{32} =   t_4 r_1 r_3 e^{-i \Phi_4} \nonumber \\
U_{33} =  (r_4 r_5 - t_3 t_4 t_5) e^{-i \Phi_4}  \quad
&U_{34} =  -(t_5 r_4 + t_3 t_4 r_5) e^{-i \Phi_4}\nonumber \\[1ex]
U_{41} =  -(t_2 r_1 + t_1 t_3 r_2) e^{-i \Phi_3}  \quad
&U_{42} =  -(t_1 t_2 - t_3 r_1 r_2) e^{-i \Phi_3} \nonumber \\
U_{43} =  t_5 r_2 r_3 e^{-i \Phi_3}  \quad
&U_{44}  =  r_2 r_3 r_5 e^{-i \Phi_3}
\end{array}
\end{equation}
which can be substituted in Eq.\ (\ref{eq-lsd-statUomega}).  Then the
energy levels, $E_k$, of the closed RG unit including phases
$\Phi_j(E)=\Phi_j(E_k)$, are the energies for which one of the four
eigenvalues of the matrix $U$ is equal to one. Thus, the calculation of
the energy levels reduces to a diagonalization of the $4\times 4$
matrix.

The crucial step now is the choice of the energy dependence
$\Phi_j(E)$.  If each loop in Fig.\ \ref{fig-lsd-RGstruct} is viewed
as a closed equipotential as it is the case for the first step of the
RG procedure \cite{ChaC88}, then $\Phi_j(E)$ is a true magnetic phase
and changes linearly with energy with a slope governed by the actual
potential profile, which, in turn, determines the drift velocity. Thus
\begin{equation}
\label{eq-lsd-PhiE}
\Phi_j(E)=\Phi_{0,j}+2\pi\frac{E}{s_j},
\end{equation}
where a random part, $\Phi_{0,j}$, is uniformly distributed within
$[0, 2\pi]$, and $2\pi/s_j$ is a random slope. Here the coefficient
$s_j$ acts as an initial level spacing connected to the loop $j$ of
the RG unit by defining a periodicity of the corresponding phase.
Strictly speaking, the dependence (\ref{eq-lsd-PhiE}) applies only for
the first RG step.  At each following step, $n>1$, $\Phi_j(E)$ is a
complicated function of $E$ which carries information about all energy
scales at previous steps. However, in the spirit of the RG approach,
one can assume that $\Phi_j(E)$ can still be linearized within a relevant
energy interval.  The conventional RG approach suggests that different
scales in {\em real} space can be decoupled.  Linearization of
Eq.\ (\ref{eq-lsd-PhiE}) implies a similar decoupling in {\em
  energy} space.

\begin{figure}
\centerline{\includegraphics[width=0.48\columnwidth]{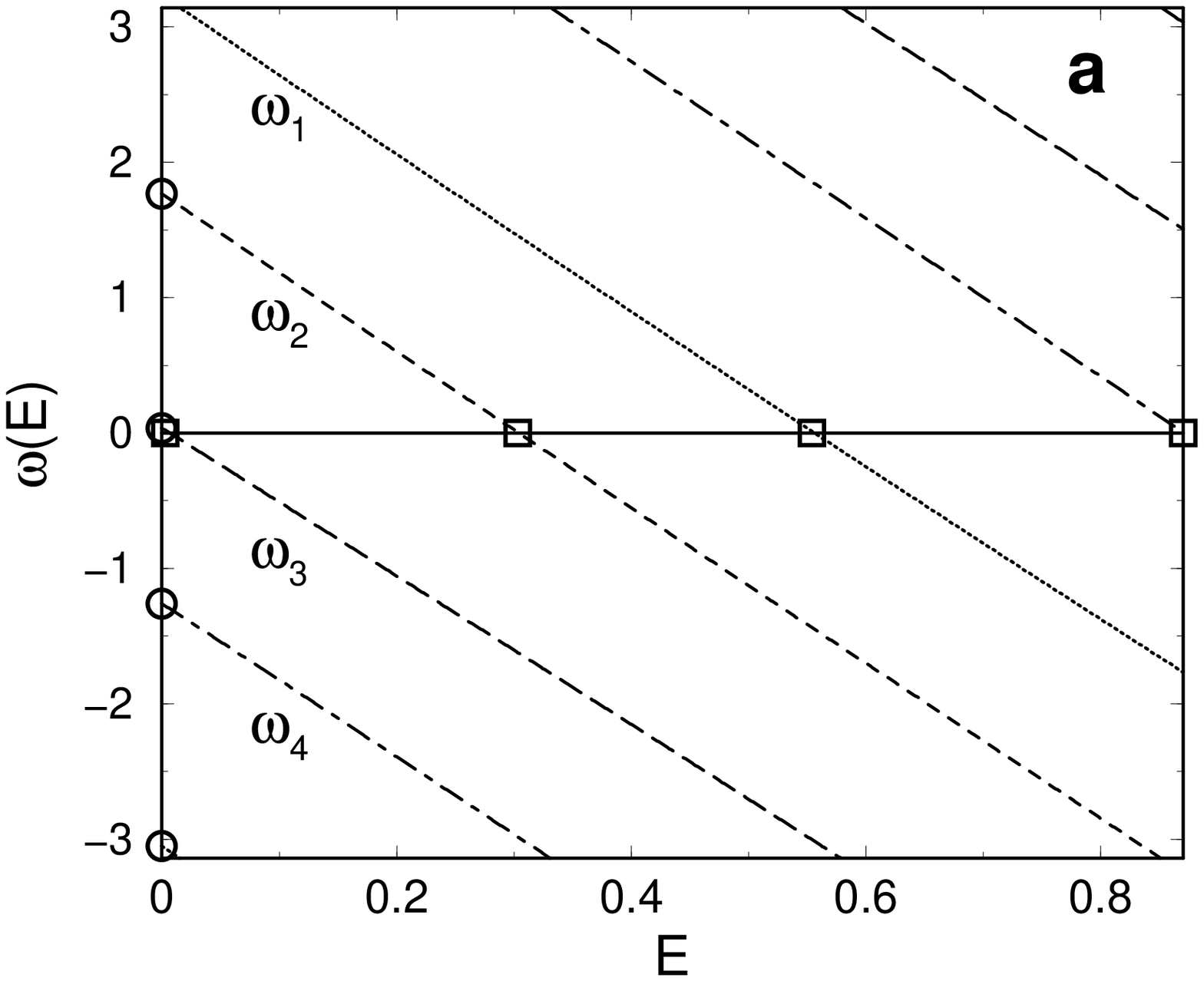}
\includegraphics[width=0.48\columnwidth]{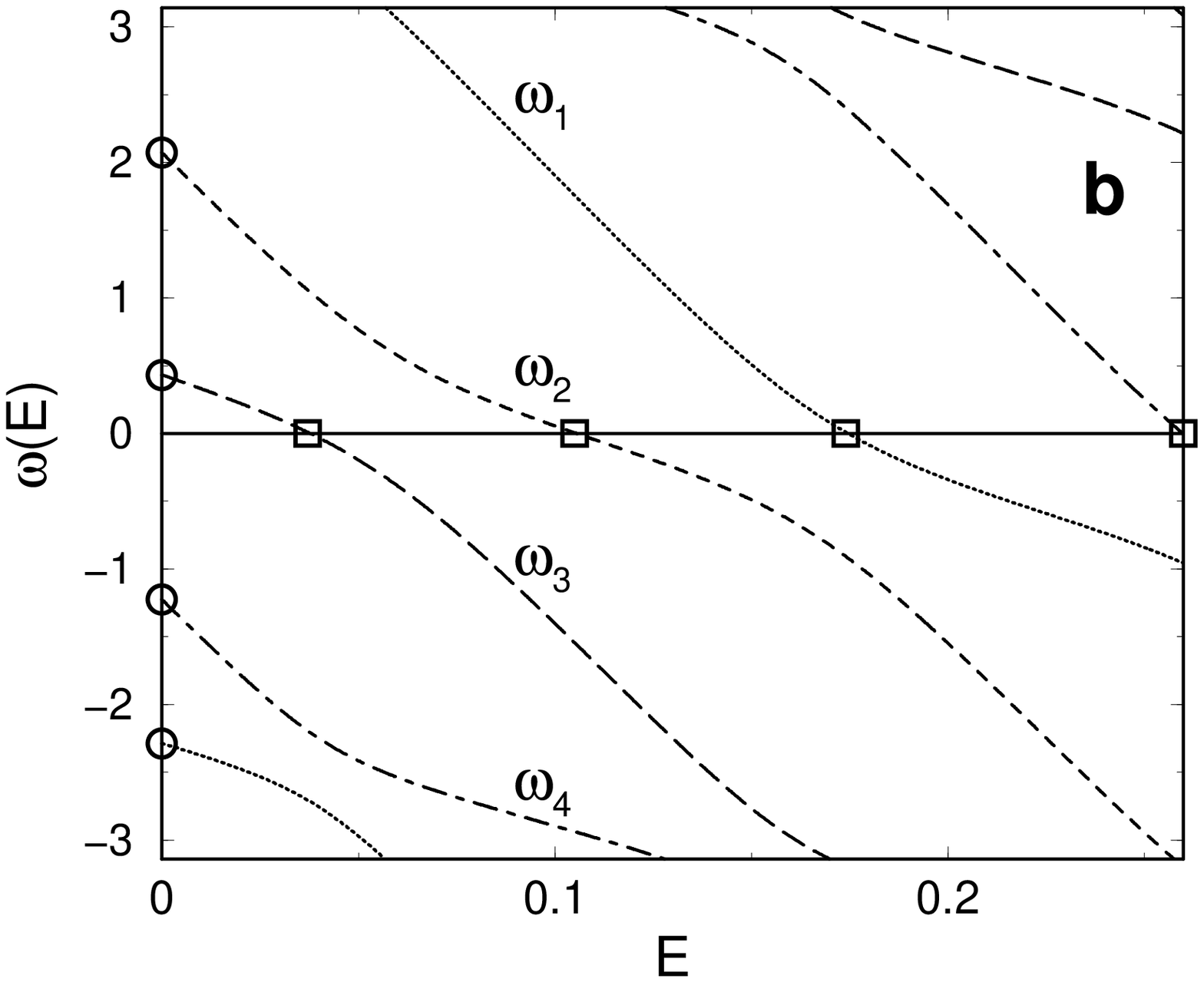}}
\caption{\label{fig-lsd-WoE}
  Energy dependence of the quasieigenenergies $\omega$ for two sample
  configurations. Instead of using the quasispectrum obtained from
  $\omega_l(E=0)$ ($\bigcirc $) we calculate the real eigenenergies
  according to $\omega(E_k)=0$ ($\Box $). Different line styles
  distinguish different $\omega_l(E)$. We emphasize that the observed
  behavior varies from sample to sample between remarkably linear (a)
  and strongly nonlinear (b)}
\end{figure}

With $\Phi_j(E)$ given by Eq.\ (\ref{eq-lsd-PhiE}), the statistics
of energy levels determined by the matrix equation
(\ref{eq-lsd-statUomega}) is obtained by averaging over the random
initial phases $\Phi_{0,j}$ and values $t_i$ chosen randomly
according to a distribution $P(t)$.  For every realization the
levels $E_k$ are computed from the solutions $\omega(E_k)=0$ of
Eq.\ (\ref{eq-lsd-statUomega}) yielding $3$ level spacings as
illustrated in Fig.\ \ref{fig-lsd-WoE}.  Thus the situation is
comparable with estimating the true random matrix ensemble
distribution functions from small, say, $2\times 2$ matrices only
\cite{Wig51,Meh91}.
Within the RG approach, the slopes $s_j$ as in Eq.\ (\ref{eq-lsd-PhiE})
determine the level spacings at the first step.  They are randomly
distributed with a distribution function $P_0(s)$.  Subsequent averaging
over many realizations yields the ELS, $P_1(s)$, at the second step.
Then the key element of the RG procedure, as applied to the level
statistics, is using $P_1(s)$ as a {\em distribution of slopes} in Eq.\
(\ref{eq-lsd-PhiE}).  This leads to the next-step ELS and so on.

The approach of this work relies on the {\em real} eigenenergies
of the RG unit.  The simpler computation of the spectrum of
quasienergies adopted in large-scale simulations within the CC
model \cite{KleM97,Met98b} cannot be applied since the energy
dependence of phases $\Phi_j$ in the elements of the matrix is
neglected and only the random contributions, $\Phi_{0,j}$, are
kept. Nevertheless it is instructive to compare the two procedures
as presented in Fig.\ \ref{fig-lsd-WoE}.  The figure shows the
dependence of the $4$ quasienergies $\omega_k$ on the energy $E$
calculated for two single sample RG units, with $t_i$ chosen from
the critical distribution $P_{\rm c}(t)$. The energy dependence of
the phases $\Phi_j$ was chosen from the ELS of the unitary random
matrix ensemble (GUE) according to Eq.\ (\ref{eq-lsd-PhiE}).  It
is seen that the dependences $\omega(E)$ range from remarkably
linear and almost parallel (Fig.\ \ref{fig-lsd-WoE}a) to strongly
nonlinear (Fig.\ \ref{fig-lsd-WoE}b).
%

\subsection{The Shape of the ELS at the QH Transition}

First, let us turn our attention to the shape of the ELS at the QH
transition.  As starting distribution $P_0(s)$ of the RG iteration, we
choose the ELS of GUE, since previous simulations \cite{KleM97,BatS96}
indicate that the critical ELS is close to GUE.  According to $P_0(s)$,
$s_j$ is drawn randomly and $\Phi_j$, $j= 1,\ldots,4$ is set as in Eq.\
(\ref{eq-lsd-PhiE}).  For the transmission coefficients of the SP the FP
distribution $P_{\rm c}(t)$, obtained in Sect.\ \ref{sec-rg-result}, is
used as initial distribution $P_0(t)$.  And from $P_0(t)$, the $5$
$t_i$, $i=1,\ldots,5$, are selected.
As in Sect.\ \ref{sec-rg-result} the RG transformation
(\ref{eq-qhrg}) is used to compute $10^{7}$ super-transmission
coefficients $t'$.  The accumulated distribution $P_1(t')$ is
again discretized in at least $1000$ bins, such that the bin width
is typically $0.001$ for the interval $t\in [0, 1]$. $P_1(t')$ is
then smoothed by a Savitzky-Golay filter \cite{PreFTV92} in order
to decrease statistical fluctuations. By finding solutions
$\omega(E_k)=0$ of Eq.\ (\ref{eq-lsd-statUomega}) the new ELS
$P_1(s')$ is constructed from the ``unfolded'' energy-level
spacings $s'_m=(E_{m+1}-E_{m})/\Delta$, where $m=1,2,3$,
$E_{k+1}>E_{k}$ and the mean spacing $\Delta= (E_4-E_1)/3$.  Due
to the ``unfolding'' \cite{Haa92} with $\Delta$, the average
spacing is set to one for each sample and in each RG-iteration
step spacing data of $2\times 10^6$ super-SP's can be
superimposed. The resulting ELS is discretized in bins with
largest width $0.01$.
In the following iteration step the procedure is repeated using the
$P_1$'s as initial distributions. Convergence of the iteration
process is assumed when the mean-square deviations of both
distributions $P_n(t)$ and $P_n(s)$ deviate by less than $10^{-4}$
from predecessors $P_{n-1}(t)$ and $P_{n-1}(s)$.
Once the (unstable) FP has been reached, the $P_n$'s should in principle
remain unchanged during all further RG iterations.  Our simulations show
\cite{CaiRSR01} that unavoidable numerical inaccuracies sum up within
several further iterations and lead to a drift away from the FP.  In
order to stabilize our calculation, we therefore use in every RG step
instead of $P_n(t)$ the FP distribution $P_{\rm c}(t)$.  This trick does
not alter the results but speeds up the convergence of the RG for
$P_{\rm c}(s)$ considerably.

\begin{figure}
  \centerline{\includegraphics[width=0.7\columnwidth]{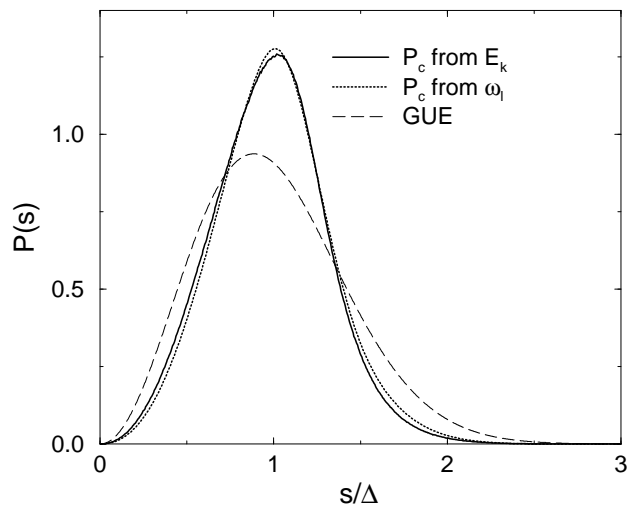}}
\caption{\label{fig-lsd-PsFP}
  FP distributions $P_{\rm c}(s)$ obtained from the spectrum of
  $\omega_l(E=0)$ and from the RG approach using the real
  eigenenergies $E_k$ in comparison to the ELS for GUE. As in all
  other graphs $P(s)$ is shown in units of the mean level spacing
  $\Delta$}
\end{figure}

This now enables us to determine the critical ELS $P_{\rm c}(s)$. The RG
iteration converges rather quickly after only $2-3$ RG steps.  The
resulting $P_{\rm c}(s)$ is shown in Fig.\ \ref{fig-lsd-PsFP} together
with the ELS for GUE.
$P_{\rm c}(s)$ exhibits the expected features, namely, level repulsion
for small $s$ and a long tail at large $s$, but the overall shape of
$P_{\rm c}(s)$ differs noticeably from GUE.  In the previous large-size
lattice simulations \cite{KleM97,BatS96} the obtained critical ELS was
much closer to GUE than $P_{\rm c}(s)$ in Fig.\ \ref{fig-lsd-PsFP}. This
fact, however, does not reflect on the accuracy of the RG approach.
Indeed, as it was demonstrated recently, the critical ELS -- although
being system size independent --- nevertheless depends on the geometry
of the samples \cite{PotS98} and on the specific choice of boundary
conditions \cite{BraMP98,SchP98}.  Sensitivity to the boundary
conditions does not affect the asymptotics of the critical distribution,
but rather manifests itself in the shape of the ``body'' of the ELS.
Recall now that the boundary conditions which have been imposed to
calculate the energy levels (dashed lines in Fig.\
\ref{fig-lsd-RGstruct}) are {\em non-periodic}.

As mentioned above, the critical ELS has also been computed previously
by diagonalizing $U(0)$ and studying the distribution of {\em
  quasienergies}. In Fig.\ \ref{fig-lsd-PsFP} the result of this
procedure using the present RG approach is shown.  It appears that the
resulting distribution is almost {\em identical} to $P_{\rm c}(s)$.
This observation is highly non-trivial, since, as follows from Fig.\
\ref{fig-lsd-WoE}, there is no simple relation between the energies and
quasienergies.
%

\subsection{Small and Large $s$ Behavior}

\begin{figure}[t]
  \centerline{\includegraphics[width=0.48\columnwidth]{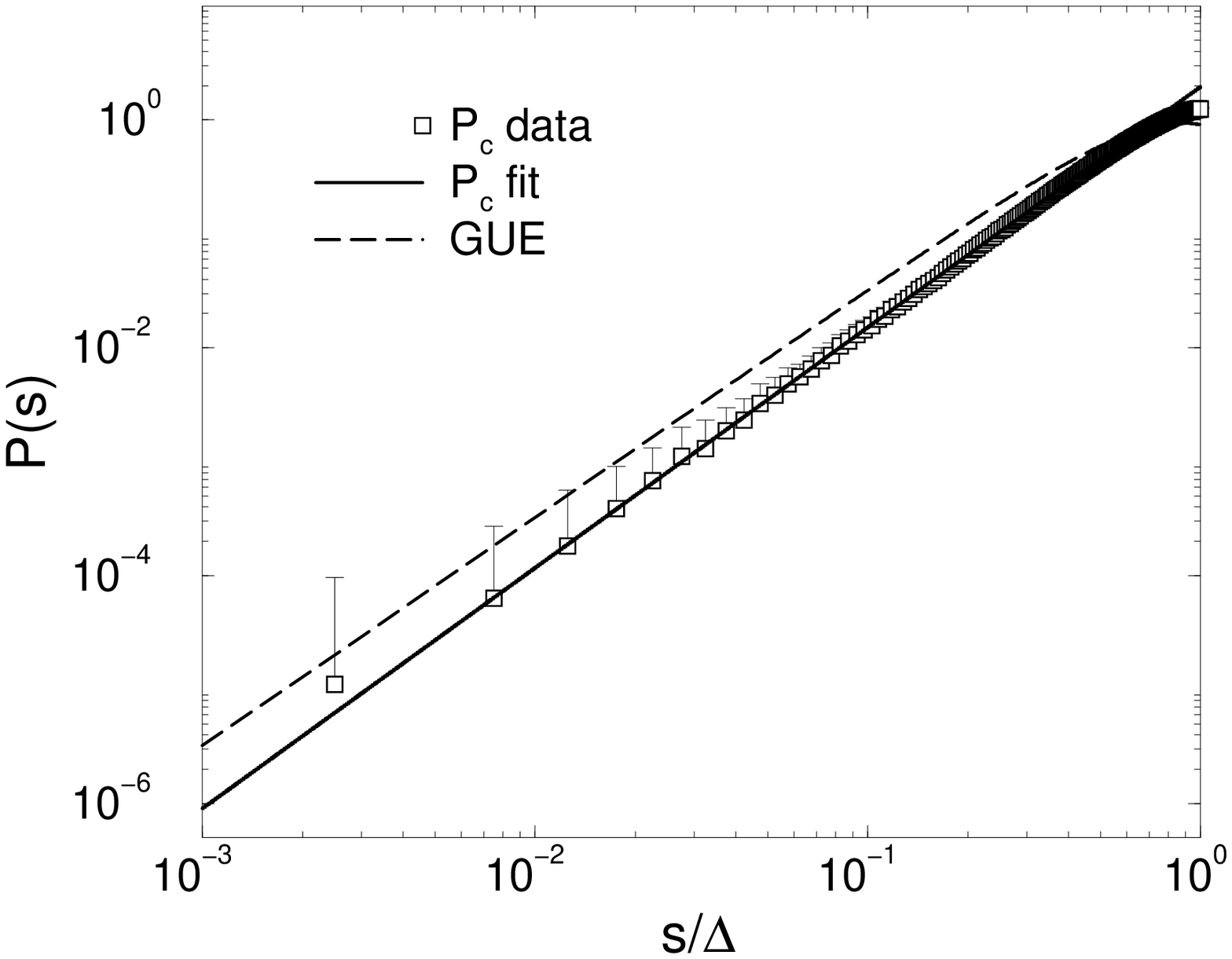}
\includegraphics[width=0.48\columnwidth]{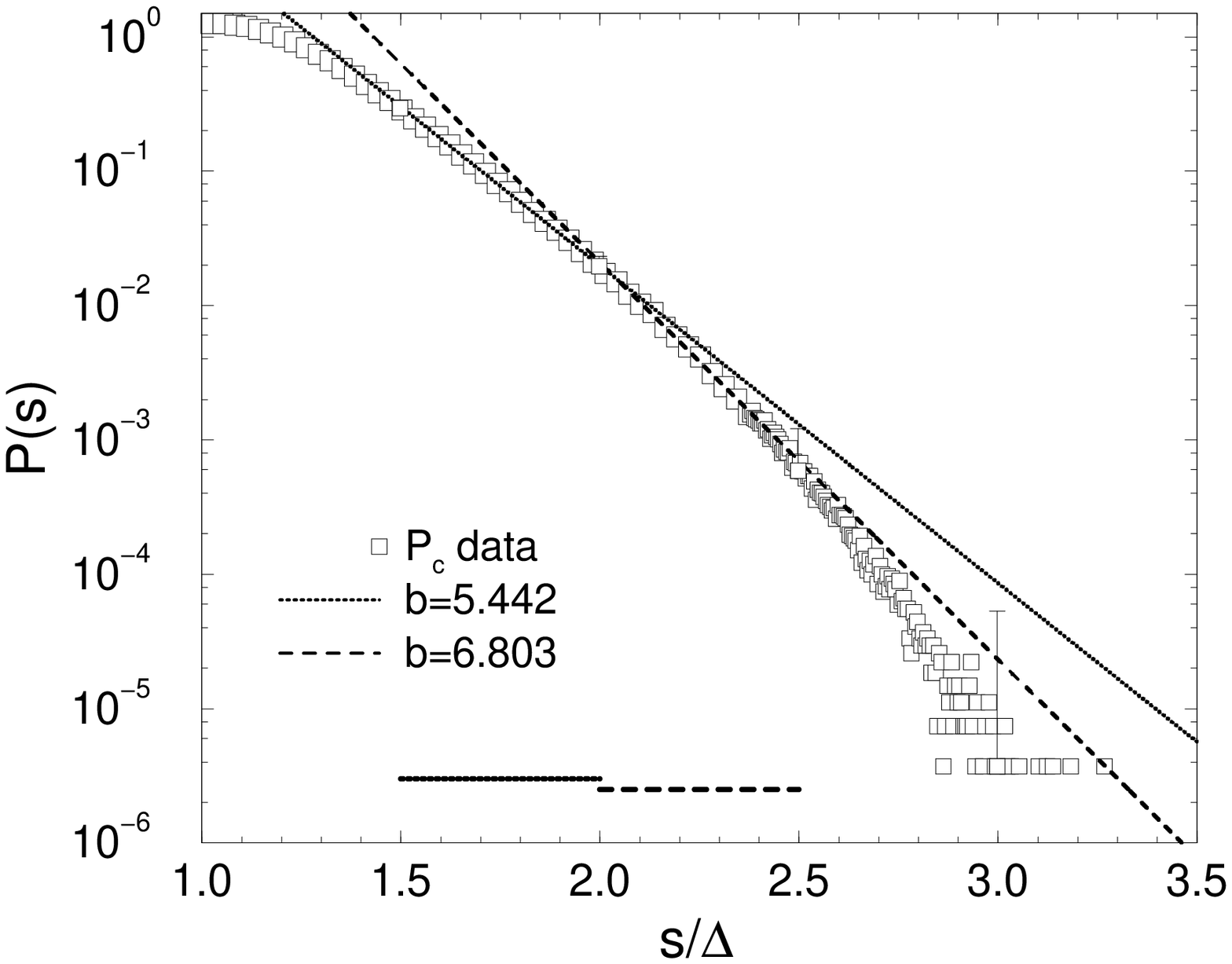}}
\caption{\label{fig-lsd-PsFPsmall}
Left:
  FP $P_{\rm c}(s)$ for small $s$ in agreement with the predicted
  $s^2$ behavior. Due to the log-log plot errors are shown in the
  upper direction only.
Right:
The large $s$ tail of $P_{\rm c}(s)$ compared with fits according
  to the predictions of Ref.\ \protect\cite{ShkSSL93} (lines).
  The interval used for fitting is indicated by the bars close to the
  lower axis. For clarity errors are shown in upper direction and for
  $s/\Delta=1.5,2.0,2.5,3.0$ only.  For $s/\Delta<2.4$, only every
  $5$th data point is drawn by a symbol}
\end{figure}

As we have seen, the general shape of the critical ELS is not
universal.  However, the small-$s$ behavior of $P_{\rm c}(s)$ must be
the same as for GUE, namely $P_{\rm c}(s) \propto s^2$. This is
because delocalization at the QH transition implies level repulsion
\cite{ShkSSL93,KleM97,Met98b,BatS96,FyoM97,KawOSO96,BatSZK96,FeiAB95,OhtO95,MetV98,BatSK98,Met99,OnoOK96}.
In Fig.\ \ref{fig-lsd-PsFPsmall} we show that this is also true for
the RG approach. The given error bars of our numerical data are
standard deviations computed from a statistical average of $100$ FP
distributions each obtained for different random sets of $t_i$'s and
$\Phi_j$'s within the RG unit. In general, within the RG approach, the
$s^2$-asymptotics of $P(s)$ is most natural.  This is because the
levels are found from diagonalization of the $4\times 4$ unitary
matrix with absolute values of elements widely distributed between $0$
and $1$.
The right form of the large-$s$ tail of $P(s)$ is Poissonian, $P_{\rm
  c}(s)\propto \exp(-bs)$ \cite{ShkSSL93} for $s \geq 3\Delta$.  The
data has a high accuracy only for $s/\Delta \leq 2.5$. For such $s$, the
distribution $P_{\rm c}(s)$ does not yet reach its large-$s$ tail and
the fit parameters shown in Fig.\ \ref{fig-lsd-PsFPsmall} depend largely
on the $s$-interval chosen.

\section{Scaling Results for the ELS}
\label{sec-lsd-fss}

\subsection{Finite-Size Scaling at the QH Transition}

The critical exponent $\nu$ of the QH transition governs the
divergence of the correlation length $\xi_\infty$ as a function of
the control parameter $z_0$, i.e.
\begin{equation}
\label{eq-lsd-diverge}
\xi_\infty(z_0) \propto |z_0-z_{\rm c}|^{-\nu},
\end{equation}
where $z_{\rm c}$ is the critical value. For the QH transition
$\nu\approx 2.35$ has been calculated by a variety of numerical methods
\cite{ChaC88,LeeWK93,Huc92} and is in agreement with the experimental
estimates $\nu\approx 2.3$ \cite{HohZH01a,KocHKP91,HohZH02}.  As
presented in Sect.\ \ref{sec-rg-result} the RG approach for the
conductance distribution yields a rather accurate value $\nu
=2.39\pm0.01$.
In order to extract $\nu$ from the ELS the one-parameter-scaling
hypothesis \cite{AbrALR79} is employed. This approach describes the
rescaling of a quantity $\alpha(N; \{z_i\})$ --- depending on (external)
system parameters $\{z_i\}$ and the system size $N$ --- onto a single
curve by using a scaling function $f$
\begin{equation}
\label{eq-lsd-scal}
\alpha\left(N;\{z_i\}\right)=f\left(\frac{N}{\xi_\infty(\{z_i\})}\right) .
\end{equation}
Since Eq.\ (\ref{eq-lsd-diverge}), as indicated by ``$\infty$'', holds
only in the limit of infinite system size, we now use the scaling
assumption to extrapolate $f$ to $N\rightarrow\infty$ from the
finite-size results of the computations.  The knowledge about $f$ and
$\xi_\infty$ then allows to derive the value of $\nu$.

We use the natural parametrization $t=(e^z+1)^{-1/2}$ \cite{GalR97},
such that $z$ can be identified with a dimensionless electron energy.
The universal conductance distribution at the transition, $P_{\rm
  c}(G)$, corresponds to a distribution $Q_{\rm c}(z)$ \cite{CaiRSR01}
which is symmetric with respect to $z=0$ and has a shape close to a
Gaussian.  The RG procedure for the conductance distribution converges
and yields $Q_{\rm c}(z)$ only if the initial distribution is an even
function of $z$.  This suggests to choose as a control parameter in Eq.\
(\ref{eq-lsd-scal}), the position $z_0$ of the maximum of the function
$Q(z)$.  The meaning of $z_0$ is an electron energy measured from the
center of the Landau band.  The fact that the QH transition is
infinitely sharp implies that for any $z_0\ne 0$, the RG procedure
drives the initial distribution $Q(z-z_0)$ towards an insulator, either
with complete transmission of the network nodes (for $z_0>0$) or with
complete reflection of the nodes (for $z_0<0$).

\subsection{Scaling for $\alpha_{\rm P}$ and $\alpha_{\rm I}$}

\begin{figure}
  \centerline{\includegraphics[width=0.7\columnwidth]{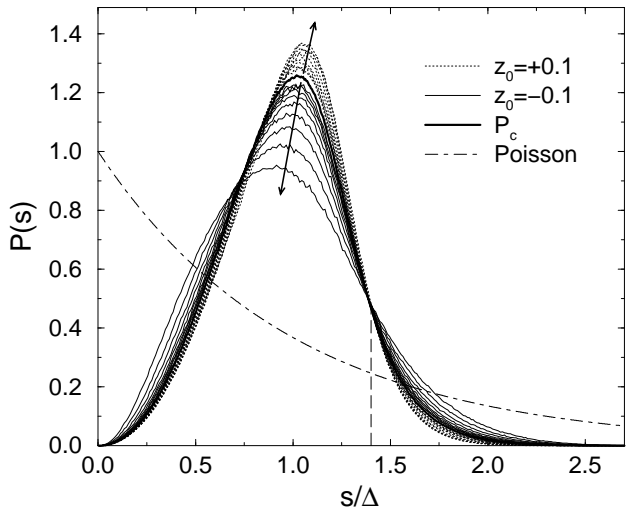}}
\caption{\label{fig-lsd-LSDshift}
  RG of the ELS used for the computation of $\nu$. The dotted lines
  corresponds to the first $9$ RG iterations with an initial
  distribution $P_0$ shifted to the metallic regime ($z_0=0.1$) while
  the thin full lines represent results for a shift toward
  localization ($z_0=-0.1$). Within the RG procedure the ELS moves
  away from the FP as indicated by the arrows. At $s/\Delta\approx
  1.4$ the curves cross at the same point -- a feature we exploit
  when deriving a scaling quantity from the ELS  }
\end{figure}

\begin{figure}
  \centerline{\includegraphics[width=0.48\columnwidth]{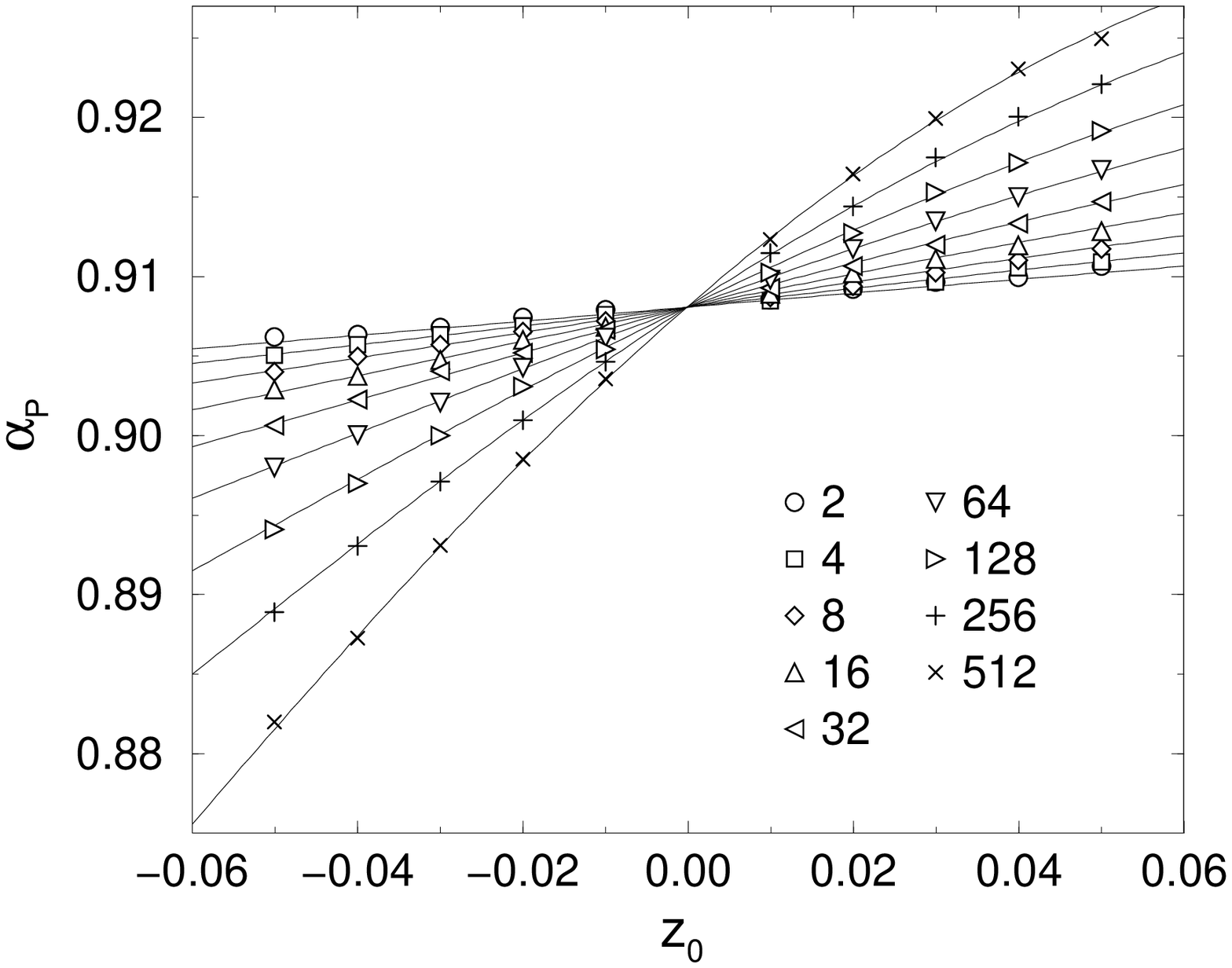}
\includegraphics[width=0.48\columnwidth]{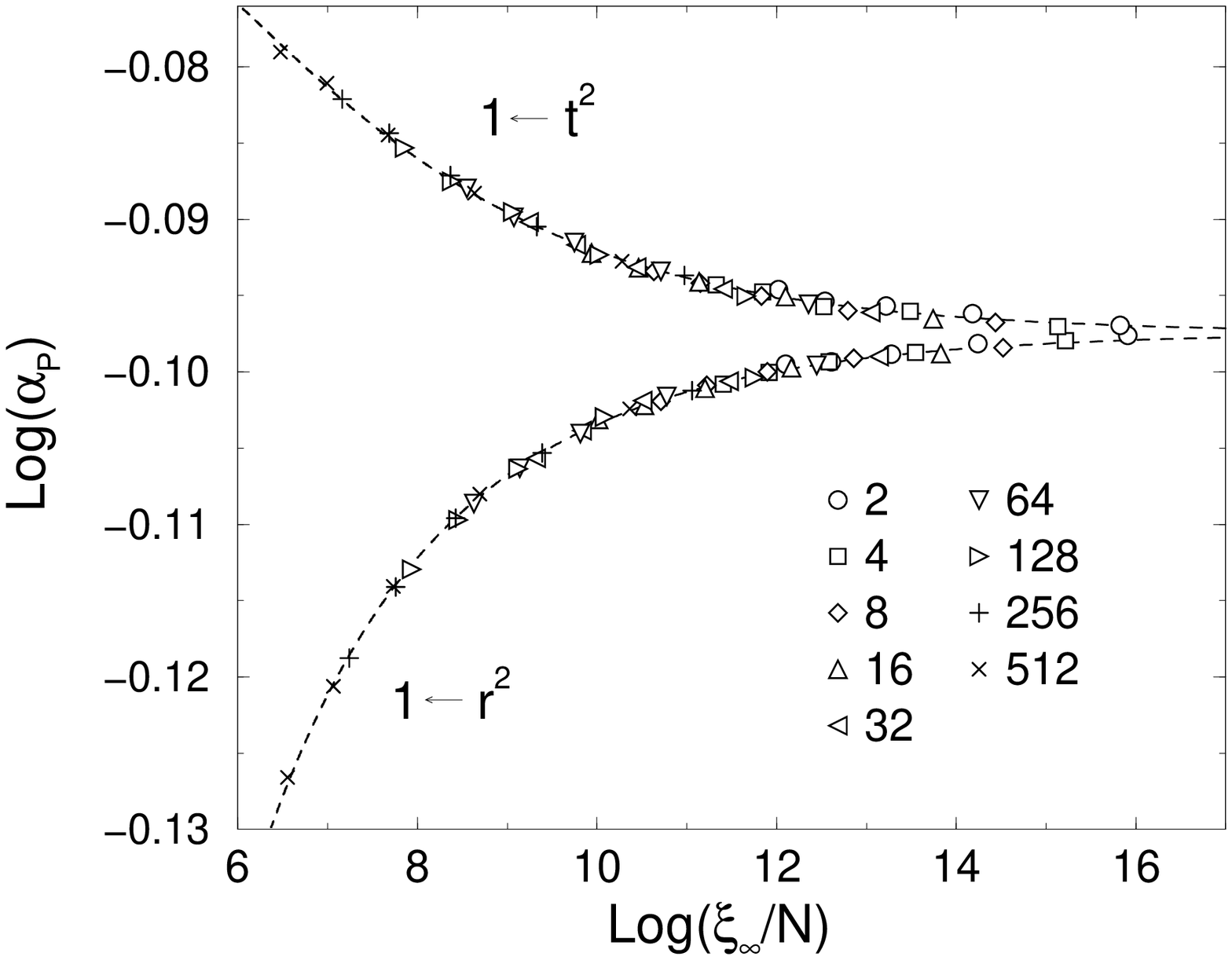}}
\caption{\label{fig-lsd-AlphaVsZ0}\label{fig-lsd-AlphaFSS}
Left:
  Behavior of $\alpha_{\rm P} $ at the QH transition as results of the
  RG of the ELS. Data are shown for RG iterations $n=1,\ldots,9$
  corresponding to effective system sizes $N=2^n=2,\ldots,512$. Full
  lines indicate the functional dependence according to FSS using the
  $\chi^2$ minimization with ${\cal O}_{\nu}=2$ and ${\cal O}_{z}=3$.
Right: 
  FSS curves resulting from the $\chi^2$ fit of our data shown in Fig.\
  \protect\ref{fig-lsd-AlphaVsZ0}.  Different symbols correspond to
  different effective system sizes $N=2^n$.  The data points collapse
  onto a single curve indicating the validity of the scaling approach
  }
\end{figure}

In principle, one is now free to choose for the finite-size scaling
analysis (FSS) any characteristic quantity $\alpha(N;z_0)$ constructed
from the ELS which has a systematic dependence on system size $N$ for
$z_0\ne0$ while being constant at the transition $z_0=0$.  Because of
the large number of possible choices
\cite{ShkSSL93,BatS96,HofS94b,ZhaK95b,ZhaK95c,ZhaK97} a restriction to
two appropriate quantities is made which are obtained by integration
of the ELS and have already been successfully used in Refs.\
\cite{HofS94b,HofS93}, namely
\begin{equation}
  \alpha_{\rm P} = \int^{s_0}_0 P(s) ds, \quad \mbox{and} \quad
  \alpha_{\rm I} = \frac{1}{s_0}\int^{s_0}_0 I(s)ds ,
\end{equation}
with $I(s)= \int^{s}_0 P(s') ds'$. The integration limit is chosen as
$s_0= 1.4$ which approximates the common crossing point \cite{HofS94b}
of all ELS curves as can be seen in Fig.\ \ref{fig-lsd-LSDshift}.
Thus $P(s_0)$ is independent of the distance $|z-z_{\rm c}|$ to the
critical point and the system size magnification $N$.  Since
$\alpha_{\rm I,P}(N,z_0)$ is analytical for finite $N$, one can expand
the scaling function $f$ at the critical point. The first order
approximation yields
\begin{equation}
  \alpha(N,z_0)\sim \alpha(N,z_{\rm c})+a |z_0-z_{\rm c}| N^{1/\nu}
\end{equation}
where $a$ is a coefficient. For our calculation we use higher
order expansions \cite{SleO99a} expanding $f$ twice, first, in
terms of Chebyshev polynomials of order ${\cal O}_\nu$ and,
second, as Taylor expansion with terms $|z_0-z_c|$ in the order
${\cal O}_z$. This procedure allows to describe deviations from
linearity in $|z_0-z_c|$ at the transition.  Contributions from an
irrelevant scaling variable can be neglected since the transition
point $z_0=0$ is known. In Fig.\ \ref{fig-lsd-AlphaVsZ0} the
resulting fits for $\alpha_{\rm P}$ at the transition are shown.

The fits are chosen in a way such that the total number of
parameters is kept at a minimal value and the fit agrees well with
the numerical data. The corresponding scaling curves for
$\alpha_{\rm P}$ are displayed in Fig.\ \ref{fig-lsd-AlphaFSS}. In
the plots the two branches for complete reflection ($z_0<0$) and
complete transmission ($z_0>0$) can be distinguished clearly.  In
order to estimate the error of the fitting procedure the results
for $\nu$ obtained by different orders ${\cal
  O}_\nu$ and ${\cal O}_z$ of the expansion, system sizes $N$, and
regions around the transition are compared.  A part of the over $100$
fit results together with the standard deviation of the fit are given in
Ref.\ \cite{CaiRR03}.  The value of $\nu$ is calculated as average of all
individual fits where the resulting error of $\nu$ was smaller than
$0.02$ resulting in $\nu=2.37\pm0.02$. This is in excellent agreement
with the previously quoted results \cite{ChaC88,LeeWK93,Huc92,CaiRSR01}.

\section{Test of Different SP Unit}
\label{sec-rg-diff}
Apparently, the quality of the RG approach crucially depends on
the choice the RG unit. For the construction of a proper chosen RG
unit two conflicting aspects have to be considered. (i) With the
size of the RG unit also the accuracy of the RG approach increases
since the RG unit can preserve more of the connectivity of the
original network. (ii) As a consequence of larger RG units the
computational effort for solving the scattering problem rises,
especially in the case where an analytic solution, as Eq.\
(\ref{eq-qhrg}), is not attained.  Because of these reasons
building an RG unit is an optimization problem depending mainly on
the computational resources available.  As mentioned in the
previous Section larger RG units were already studied in
\cite{WeyJ98,JanMMW98,JanMW98}. In these works the authors could
not benefit from an analytic solution and achieve only a similar
and less accurate statistics in comparison with the results
presented here.  In this Section the opposite case is studied
using a small RG unit proposed in Ref.\ \cite{ZulS01} in the
context of the Hall resistivity.

\begin{figure}[t]
\begin{center}
\includegraphics[width=0.4\columnwidth]{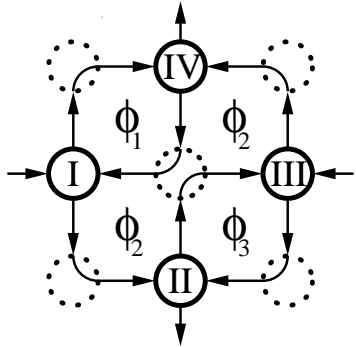}
\caption{\label{fig-rg-RGstruct4SP}
  RG unit constructed from $4$ SP's indicated by full circles.  Some
  connectivity is neglected (dotted circles). The phases $\Phi_j$ are
  accumulated by the electron motion (arrows) along contours of the
  energy potential }
\end{center}
\end{figure}
The super-SP now consists only of $4$ SP's as shown in Fig.\
\ref{fig-rg-RGstruct4SP}.  It resembles the $5$ SP's unit (Fig.\
\ref{fig-RGstruct}) used previously leaving out the SP in the
middle of the structure. Again the scattering problem can be
formulated as a system of now $8$ equations which is solved
analytically
\begin{equation}
\label{eq-rg-qhrg_4SP}
t'_{\rm 4 SP}=\left|\frac{  t_2 t_3 e^{i \Phi_2}(r_1 r_4 e^{-i \Phi_1} - 1) + t_1 t_4 (r_2 r_3 e^{i \Phi_3} - 1)}
{ (1 -r_2 r_3  e^{i \Phi_3})(1 - r_1 r_4 e^{i \Phi_1}) +  t_1 t_2 t_3 t_4 e^{i \Phi_2}} \right| \quad.
\end{equation}
The result can be verified using Eq.\ (\ref{eq-qhrg}) after setting
$t_3=0$ and $r_3=1$, joining the phases $\Phi_1$ and $\Phi_4$ and
renumbering the indices.

The RG transformation (\ref{eq-rg-qhrg_4SP}) is then applied
within the  RG approach analogously to the $5$ SP unit.
First the FP distribution $P_{\rm c}(G)$ is obtained.  A comparison of
$P_{\rm c}(G)$ for both RG units is shown in Fig.\ \ref{fig-rg-pgcomp}.
In the case of the $5$ SP unit the FP distribution $P_{\rm c}(G)$ exhibits
a flat minimum around $G=0.5$, and sharp peaks close to $G=0$ and
$G=1$.  It is symmetric with respect to $G\approx 0.5$.  The $4$ SP
unit yields differing results. While $P_{\rm c}(G)$ is still rather
flat it is clearly asymmetric, which already indicates that the $4$ SP
unit can not describe all of the underlying symmetry of the CC
network.

\begin{figure}[t]
\begin{center}
\includegraphics[width=0.7\columnwidth]{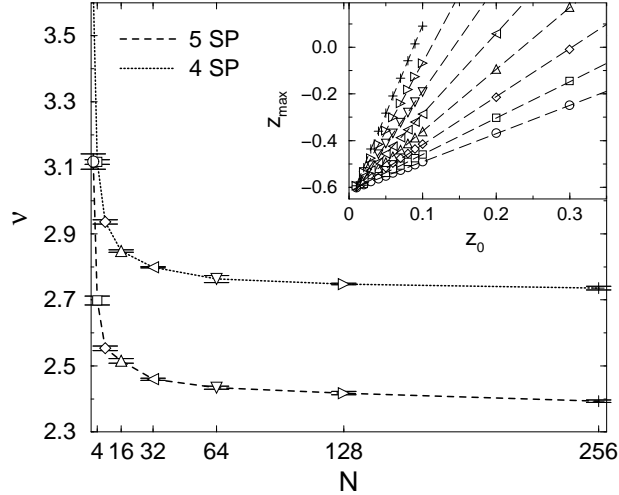}
\caption{\label{fig-rg-nucomp}
  The critical exponent $\nu$ as function of the effective system size
  $N=2^n$ for $4$ SP (dotted line) and $5$ SP unit (dashed line).
  Inset: Maximum $z_{\rm max}$ of $Q(z)$ vs.\ initial shift $z_0$ for $8$ RG
  iterations (symbols) using $4$ SP. Dashed lines indicate linear
  fits}
\end{center}
\end{figure}

The $P_{\rm c}(G)$ for the $4$ SP unit is then used in the calculation
of the critical exponent $\nu$ to construct the shifted initial
distributions $Q_0(z)$.  The behavior of $\nu$ as function of $n$ for
the $4$ and $5$ SP RG units is demonstrates in Fig.\
\ref{fig-rg-nucomp}.  Both curves approach convergence monotonously
from larger values of $\nu$. During all iteration steps, $\nu$ for the
$4$ SP differs from the $5$ SP result by an almost constant positive
shift.  After $8$ iterations, which equals an increase of system size
by a factor of $256$, one finds $\nu_{\rm 5 SP}=2.39\pm0.01$ and
$\nu_{\rm 4 SP}=2.74\pm0.02$.  The error describes a confidence
interval of $95\%$ as obtained from the fit to a linear behavior.  The
result for $\nu_{\rm 4 SP}$ deviates clearly from the five
SP result and also from the values obtained by other
methods \cite{ChaC88,LeeWK93,Huc92}.
%
%
In addition to these findings also the discussion in Sect.\
\ref{sec-rg-result} indicates that the $4$ SP RG unit fails to
describe the critical properties at the QH transition correctly. This
fact underlines again the importance of the RG unit for a successful
application of the RG approach.

\section{Conclusions}

The version of the network model \cite{Sha82} that has been most widely
studied in the context of the QH effect, is the CC model \cite{ChaC88},
describing the electron motion in a disordered system in a strong
magnetic field limit. The fact that the RG approach, within which the
correlations between different scales are neglected, describes the
results of the large-scale simulations of the CC model so accurately,
indicates that only a few spatial correlations within each scale are
responsible for the critical characteristics of the quantum Hall
transition. More precisely, the structure of the eigenstates of a
macroscopic sample at the transition can be predicted from the analysis
of a single RG unit consisting of only five nodes. Further applications
of this approach to the computation of the Hall resistance and the
plateau-to-insulator transition shall be published elsewhere.

We thank B.\ Huckestein, M.E.\ Raikh, M.\ Schreiber, and U.\
Z\"{u}licke for stimulating discussions. This work was supported
by the DFG within SFB393 and the priority research program on
quantum Hall systems. Further support was provided by a DAAD-NSF
collaborative research grant INT-0003710.


%
\clearpage
\addcontentsline{toc}{section}{Index}
\flushbottom
\printindex

\end{document}